\documentclass[twocolumn,english,aps,english,showpacs,preprintnumbers,groupedaddress,amsmath,amssymb,pra]{revtex4}
\usepackage[T1]{fontenc}
\usepackage[latin9]{inputenc}
\usepackage{amsmath}
\usepackage{amssymb}
\usepackage{graphicx}
\usepackage{esint}

\makeatletter
\@ifundefined{textcolor}{}
{%
 \definecolor{BLACK}{gray}{0}
 \definecolor{WHITE}{gray}{1}
 \definecolor{RED}{rgb}{1,0,0}
 \definecolor{GREEN}{rgb}{0,1,0}
 \definecolor{BLUE}{rgb}{0,0,1}
 \definecolor{CYAN}{cmyk}{1,0,0,0}
 \definecolor{MAGENTA}{cmyk}{0,1,0,0}
 \definecolor{YELLOW}{cmyk}{0,0,1,0}
}

\usepackage{babel}

\makeatother

\usepackage{babel}
\begin{document}

\title{\textmd{Simultaneous slow and fast light involving the Faraday effect}}

\author{Bruno Macke}

\author{Bernard S\'{e}gard}

\email{bernard.segard@univ-lille-1.fr}

\affiliation{Laboratoire de Physique des Lasers, Atomes et Mol\'{e}cules , CNRS et
Universit\'{e} de Lille, 59655 Villeneuve d'Ascq, France}

\date{\today}
\begin{abstract}
We theoretically study the linear transmission of linearly polarized
light pulses in an ensemble of cold atoms submitted to a static magnetic
field parallel to the direction of propagation. The carrier frequency
of the incident pulses coincides with a resonance frequency of the
atoms. The transmitted light, the electric field of which is transversal,
is examined in the polarizations parallel and perpendicular to that
of the incident pulses. We give explicit analytic expressions for the
transfer functions of the system for both polarizations and for the
corresponding group delays. We demonstrate that slow light can be
observed in a polarization, whereas fast light is simultaneously observed
in the perpendicular polarization. Moreover, we point out that, due
to the polarization post selection, the system is not necessarily
minimum phase shift. Slow light can then be obtained in situations
where an irrelevant application of the Kramers-Kronig relations could lead one
to expect fast light. When the incident light is step modulated, we
finally show that, in suitable conditions, the system enables one
to separate optical precursor and main field. 
\end{abstract}

\pacs{42.25.Bs, 42.25.Lc, 42.50.Md}

\maketitle
Dilute atomic or molecular media are precious tools for the study
of the propagation of light in material \cite{al87,si14} and, more
specifically, of the phenomena of slow light, fast light and optical
precursors \cite{bo02,mi04,ak10,che13}. With their refractive index $n$
being very close to unity, the parasitic reflections at the input
and the output of the medium (``etalon effects\textquotedblright )
that may complicate the analysis of the transmitted signals \cite{chu82}
are practically eliminated. On the other hand, the narrowness of their
absorption or gain lines originates the singular group velocities
$v_{g}$ required to observe significant slow light ($0<v_{g}\ll c,$
where $c$ is the light velocity in vacuum) or fast light ($v_{g}>c$
or $v_{g}<0$) \cite{bo02}. These group velocities are often obtained
when the carrier frequency of the probe pulses coincides with a well-marked
peak in the medium transmission in the slow-light case or a well-marked
dip in the fast-light case. In most of the experiments, these conditions
are created by applying extra fields (pump and/or coupling fields)
whose interaction with the medium is nonlinear \cite{bo02,mi04,ak10,gla12}.
It is, however, worth recalling that the pioneering experimental demonstrations
of slow and fast light have been performed without any pump or coupling
fields \cite{gr73,chu82,se85}. More recent experiments in ``natural\textquotedblright{}
atomic media are reported in \cite{ta03,ca06,ca07,si09,ke12}. The
challenge in all these experiments is to obtain significant effects
with moderate pulse distortion. 
\begin{figure}[h]
\begin{centering}
\includegraphics[width=0.95\columnwidth]{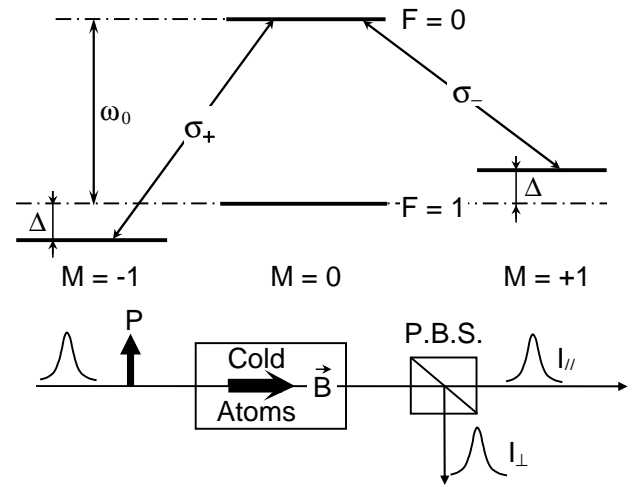} 
\par\end{centering}

\caption{Proposed level arrangement (top) and experimental setup (bottom).
$\protect\overrightarrow{B}$ is the magnetic field parallel to the
direction of the light propagation, P and PBS respectively designate
a linear polarizer and a polarization beam splitter separating the
polarizations parallel ($\parallel$) and perpendicular ($\perp$)
to the polarization of the incident pulse\textbf{.}\label{fig:SetUp}}
\end{figure}

Fast-light and slow-light experiments have also been performed in
nondispersive media by exploiting the effects of field polarization \cite{so03,so04,bru04,ha08}. In these experiments, a medium \cite{so03,so04,ha08} or fiber \cite{bru04}
with linear birefringence is placed between two linear polarizers. Note that fast light and slow light are then observed in different experimental conditions, the transition from fast to slow light being achieved by changing the orientation of the output polarizer. A similar behavior is obtained in the related system considered in \cite{bi08}.

In the present article, we propose a system combining effects of light polarization and of medium dispersion. It involves two detection channels, enabling one to observe fast light and slow light in the same experiment.
The medium consists of a cloud of cold atoms submitted to a uniform \cite{re00}, static magnetic field parallel to the direction
of propagation of the (unique) probe field (Fig. \ref{fig:SetUp}).
The incident field is transversal and linearly polarized by a polarizer
(P) and the field transmitted by the medium is received on a polarization
beam splitter (PBS) that separates the polarizations parallel and
perpendicular to that of the incident field, with both directions
of polarization being perpendicular to the direction of propagation.
Cold atoms avoid the complications of Doppler broadening and ensure
significant Faraday effects (circular birefringence and dichroism)
for moderate values of the magnetic field \cite{la01}. For simplicity,
the carrier-frequency $\omega_{c}$ of the incident pulse is assumed
to coincide with the frequency $\omega_{0}$ of the transition from
a ground level of total angular momentum $F=1$ to an excited state
$F=0$. In the presence of magnetic field the transition is split
in two components of frequency $\omega_{0}\pm\Delta$ associated with
the circular polarizations $\sigma_{\pm}$ \cite{bu02}. In a frame
rotating at the angular frequency $\omega_{0}$, the transfer functions
$H_{\pm}(\Omega)$ relating the Fourier transforms of the envelopes
of the incident and transmitted pulses for the polarizations $\sigma_{\pm}$
read  \cite{ma03,pa87} 
\begin{equation}
H_{\pm}(\Omega)=\exp\left(-\frac{\alpha\ell\gamma}{2\left(\gamma+i\Omega\mp i\Delta\right)}\right),\label{eq:1}
\end{equation}
where $\alpha$ is the resonance absorption coefficient \emph{for
the intensity} of the medium, $\ell$ is the medium thickness and
$\gamma$ is the half width at half maximum (HWHM) of the resonances.
Notice that Eq. (\ref{eq:1}) is obtained by using a time retarded
by the transit time at the velocity $c$. Decomposing the unit vector
of the linearly polarized incident field in the complex unit vectors
associated with the circular polarizations $\sigma_{\pm}$, we apply
the transfer functions $H_{\pm}(\Omega)$ to these polarizations and
project each of the resulting fields on the directions parallel and
perpendicular to that of the incident field. In this way we get the
transfer functions $H_{\parallel}(\Omega)$ and $H_{\bot}(\Omega)$
corresponding to the fields transmitted in each polarization. They
read 
\begin{equation}
H_{\parallel}(\Omega)=\frac{1}{2}\left[H_{+}(\Omega)+H_{-}(\Omega)\right],\label{eq:2}
\end{equation}
\begin{equation}
H_{\bot}(\Omega)=\frac{i}{2}\left[H_{+}(\Omega)-H_{-}(\Omega)\right].\label{eq:3}
\end{equation}
When the probe wave is a continuous wave (cw) of optical frequency
$\omega_{0}$ ($\Omega=0$), the amplitude transmissions are reduced
to $H_{\parallel}(0)=e^{-\gamma\theta/\Delta}\cos\theta$
and $H_{\bot}(0)=e^{-\gamma\theta/\Delta}\sin\theta$ where
\begin{equation}
\theta=\frac{\alpha\ell\Delta}{2\gamma\left(1+\Delta^{2}/\gamma^{2}\right)}\label{eq:4}
\end{equation}
is nothing more than the Faraday rotation angle of the field polarization
in the medium.

The study of the time-dependent regime is greatly simplified by remarking
that the transfer functions given in Eqs. (\ref{eq:2}) and (\ref{eq:3}) are
such that $H(\Omega)=H^{*}(-\Omega)$, where the asterisk stands for the
complex conjugate. Assuming that the envelope $x(t)$ of the incident
pulse is real positive (amplitude modulation), the envelope $y(t)$
of the transmitted pulse will be also real, and a simple application
of the moment theorem \cite{pa87} shows that its center of gravity
is delayed by the group delay 
\begin{equation}
\tau_{g}=\frac{i}{H(0)}\frac{dH(\Omega)}{d\Omega}\mid_{\Omega=0}=-\frac{d\Phi(\Omega)}{d\Omega}\mid_{\Omega=0},\label{eq:5}
\end{equation}
where $\Phi(\Omega)$ is the argument of $H(\Omega)$ \cite{re0}.
It also results from the relation $H(\Omega)=H^{*}(-\Omega)$ that
$\left|H(\Omega)\right|$ and $d\Phi(\Omega)/d\Omega$ are stationary
around $\Omega=0$. If the spectrum of the incident pulse is narrow
enough, the envelope $y(t)$ of the transmitted pulse can thus be
determined by approximating $H(\Omega)$ by $H(0)\exp\left(-i\Omega\tau_{g}\right)$
and reads as 
\begin{equation}
y(t)\approx H(0)\,x(t-\tau_{g}).\label{eq:6}
\end{equation}
The incident pulse will then be transmitted without significant distortion.
Keeping in mind that we use retarded times, the transmission will
be superluminal (or even have absolute time advancement) when $\tau_{g}<0$
and subluminal when $\tau_{g}>0$. 

From Eqs.(\ref {eq:1})-(\ref{eq:3}) and (\ref{eq:5}), we easily derive
the group delays for the parallel and perpendicular polarizations.
They respectively read 
\begin{equation}
\tau_{g\parallel}=-\frac{\theta}{\gamma\left(1+\Delta^{2}/\gamma^{2}\right)}\left[\left(1-\Delta^{2}/\gamma^{2}\right)+\frac{2\Delta\tan\theta}{\gamma}\right],\label{eq:7}
\end{equation}
\begin{equation}
\tau_{g\perp}=-\frac{\theta}{\gamma\left(1+\Delta^{2}/\gamma^{2}\right)}\left[\left(1-\Delta^{2}/\gamma^{2}\right)-\frac{2\Delta}{\gamma\tan\theta}\right].\label{eq:8}
\end{equation}
The previous results are valid for arbitrary values of $\Delta$.
We restrict ourselves in the following the analysis to the particular case $\Delta=\gamma$,
a condition easily met with cold atoms for reasonable magnetic fields
\cite{la01}. This case is of special interest because the group velocities
for the circular polarizations $\sigma_{+}$ and $\sigma_{-}$ are
both equal to $c$ (luminal propagation) at the carrier-frequency
of the incident pulse \cite{re1}. We then get $\theta=\alpha\ell/4$,
$H_{\parallel}(0)=e^{-\theta}\cos\theta$, $H_{\perp}(0)=e^{-\theta}\sin\theta$,
$\gamma\tau_{g\parallel}=-\theta\tan\theta$, and $\gamma\tau_{g\perp}=\theta/\tan\theta$
. A remarkable result is obtained when $\theta=\pi/4$ ($\alpha\ell=\pi$),
for which $\left|H_{\parallel}(0)\right|=\left|H_{\perp}(0)\right|=e^{-\pi/4}/\sqrt{2}$,
$\tau_{g\parallel}=-\pi/(4\gamma)$ and $\tau_{g\perp}=\pi/(4\gamma)$.
\begin{figure}[h]
\begin{centering}
\includegraphics[width=0.9\columnwidth]{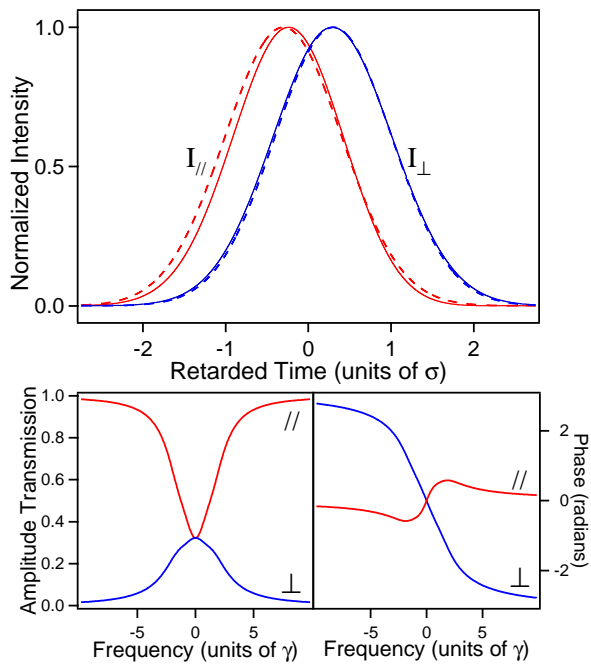} 
\par\end{centering}

\caption{Top: normalized intensity profiles of the transmitted
pulses in the polarizations $\parallel$ and $\perp$ for an incident
pulse of Gaussian envelope $x(t)=\exp\left[-t^{2}/(2\sigma^{2})\right]$.
The solid {(}dashed{)} lines are the exact profiles obtained by Fourier
transform {(}the approximate analytical profiles given by Eq.(\ref{eq:6}){)}.
Parameters: $\Delta=\gamma$, $\theta=\pi/4$ ($\alpha\ell=\pi$),
and $\sigma=2.6/\gamma$. Bottom: corresponding amplitude transmission
and phase as a function of $\Omega$.\label{fig:AlegalPI}}
\end{figure}
Figure \ref{fig:AlegalPI} shows the normalized intensity profiles
of the transmitted pulses for an incident Gaussian pulse of envelope
$x(t)=\exp\left(-t^{2}/(2\sigma^{2})\right)$ with $\sigma=2.6/\gamma$,
a value conciliating moderate pulse distortion with significant fractional
time advancement or delay. As expected, there is advancement for the
polarization parallel and delay for the polarization perpendicular,
these quantities both being nearly equal to $\pi/(4\gamma)$. Figure
 \ref{fig:AlegalPI} also shows the corresponding amplitude transmission
$\left|H(\Omega)\right|$ and phase $\Phi(\Omega)$ as functions of
$\Omega$. We see that fast light and slow light obtained in the parallel
and perpendicular polarizations are, respectively, associated with a
minimum and a maximum of transmission at $\Omega=0$. This result
is often considered as a consequence of the Kramers-Kronig relations
from which one gets 
\begin{equation}
\Phi(\Omega)=\Phi_{KK}(\Omega)=-\mathcal{H}\left\{ \ln\left|\textrm{H(}\textrm{\ensuremath{\Omega}}\textrm{)}\right|\right\},\label{eq:9}
\end{equation}
where $\mathcal{H}$ designates the Hilbert transform \cite{pa87}.
The group delay then reads
\begin{equation}
\tau_{g}=\tau_{KK}=-\frac{d\Phi_{KK}(\Omega)}{d\Omega}\mid_{\Omega=0}\label{eq:10}
\end{equation}

\begin{figure}[h]
\begin{centering}
\includegraphics[width=0.9\columnwidth]{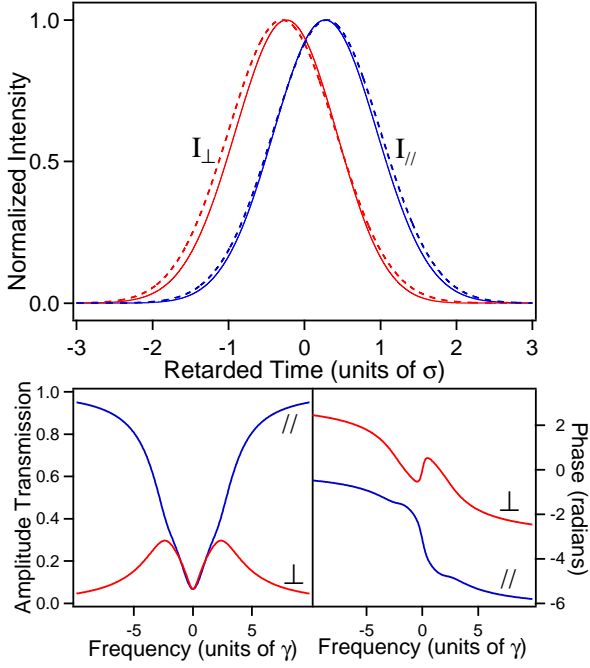} 
\par\end{centering}

\caption{Same as Fig. \ref{fig:AlegalPI} for $\Delta=\gamma$,
$\theta=3\pi/4$ ($\alpha\ell=3\pi$) and $\sigma=8.0/\gamma$. The
duration $\sigma$ of the incident pulse has been chosen in order
that the distortion of the transmitted pulse does not exceed that
of Fig. \ref{fig:AlegalPI}. \label{fig:Alegal3Pi}}
\end{figure}

In reality, Eq. (\ref{eq:9}) only holds if $H(\Omega)$ is a minimum-phase-shift
(MPS) function \cite{pa87,to56}. This condition is met for purely
propagative systems but may fail for systems involving polarizers
\cite{bi08,so03,bru04,ha08}. With $\Omega$ being continued in the complex
plane, the condition to ensure that $H(\Omega)$ is MPS is that all
its zeros lie in the upper half plane {[}$\mathrm{Im}(\Omega)>0${]}.
Assuming again that $\Delta=\gamma$, it is easily shown that the
zeros that may eventually have a negative imaginary part will occur
at 
\begin{equation}
\Omega=i\gamma\left(1-\sqrt{\frac{4\theta}{\left(2p-1\right)\pi}-1}\right)\label{eq:11}
\end{equation}
for $H_{\parallel}(\Omega)$ and at 
\begin{equation}
\Omega=i\gamma\left(1-\sqrt{\frac{2\theta}{p\pi}-1}\right)\label{eq:12}
\end{equation}
for $H_{\perp}(\Omega)$ , where $p$ is a positive integer. No such
zeros exist for $0<\theta<\pi/2$ and $H(\Omega)$ is MPS for both
polarizations. In the conditions of Fig. \ref{fig:AlegalPI} ($\theta=\pi/4$),
we have numerically verified that the group delays derived from Eqs. (\ref{eq:9})
and (\ref{eq:10}) are actually equal to the exact values $\pm\pi/(4\gamma)$.
When $\theta>\pi/2$, $H(\Omega)$ is not MPS, at least for one polarization.
It can then be written as the product of a MPS transfer function by
the transfer function $H_{AP}(\Omega)$ of a causal all-pass filter
of the form 
\begin{equation}
H_{AP}(\Omega)=\prod_{p}\left(\frac{\Omega-\Omega_{p}}{\Omega-\Omega_{p}^{*}}\right),\label{eq:13}
\end{equation}
where $\Omega_{p}$ designates the zeros of $H(\Omega)$ actually
lying in the lower complex half plane \cite{pa87,to56,pa88}. As shown in
Eqs.(\ref{eq:11}) and (\ref{eq:12}), these zeros are purely imaginary
and the contributions of $H_{AP}(\Omega)$ to add to the phase shift
$\Phi_{KK}(\Omega)$ and to the group delay $\tau_{KK}$ respectively
given by Eqs.(\ref{eq:9}) and (\ref{eq:10}) take the simple forms
\begin{equation}
\Phi_{AP}\left(\Omega\right)=-2\sum_{p}\tan^{-1}\left(\frac{\Omega}{i\Omega_{p}}\right),\label{eq:14}
\end{equation}
\begin{equation}
\tau_{AP}=-\frac{d\Phi_{AP}(\Omega)}{d\Omega}\mid_{\Omega=0}=2\sum_{p}\frac{1}{i\Omega_{p}}.\label{eq:15}
\end{equation}
Note that all the terms intervening in Eq. (\ref{eq:15}) are positive
and thus contribute to an increase of the group delay. When $\tau_{KK}<0$,
this contribution can even change the advancement predicted by the
Kramers-Kronig relations in a delay of comparable magnitude. Figure
 \ref{fig:Alegal3Pi}, obtained for $\theta=3\pi/4$ ($\alpha\ell=3\pi$),
illustrates such a case. We have then $\left|H_{\parallel}(0)\right|=\left|H_{\perp}(0)\right|=e^{-3\pi/4}/\sqrt{2}$,
$\tau_{g\parallel}=3\pi/(4\gamma)$ and $\tau_{g\perp}=-3\pi/(4\gamma)$.
For both polarizations, the amplitude transmission $\left|H(\Omega)\right|$
has a well-marked dip at $\Omega=0$. Equation (\ref{eq:12}) shows that
$H_{\perp}(\Omega)$ has no zeros in the lower half-plane and thus
is MPS. As usual, the corresponding transmitted pulse is advanced, and
we have again verified that $\tau_{g}=\tau_{KK}=-3\pi/(4\gamma)$.
On the other hand, $H_{\parallel}(\Omega)$ is not MPS. Equation (\ref{eq:11})
indeed shows that it has one active zero $\Omega_{1}=-i\gamma\left(\sqrt{2}-1\right)$
in the lower half plane. The associated time delay reads as $\tau_{AP}=2/(i\Omega_{1})=2/\left[\gamma\left(\sqrt{2}-1\right)\right]\approx4.828/\gamma$.
From Eqs. (\ref{eq:9}) and (\ref{eq:10}), we get in this case $\tau_{KK}=-2.472/\gamma$
and, finally, $\tau_{g\parallel}=\tau_{KK}+\tau_{AP}\approx2.356/\gamma$
in agreement with the expression $\tau_{g\parallel}=3\pi/(4\gamma)$
given above. We are actually in a case where the time delay has a
value nearly opposite to that derived by an irrelevant application
of the Kramers-Kronig relations. The transmitted pulse is then delayed
in spite of a dip in the system transmission at the pulse carrier
frequency.

\begin{figure}[h]
\begin{centering}
\includegraphics[width=0.9\columnwidth]{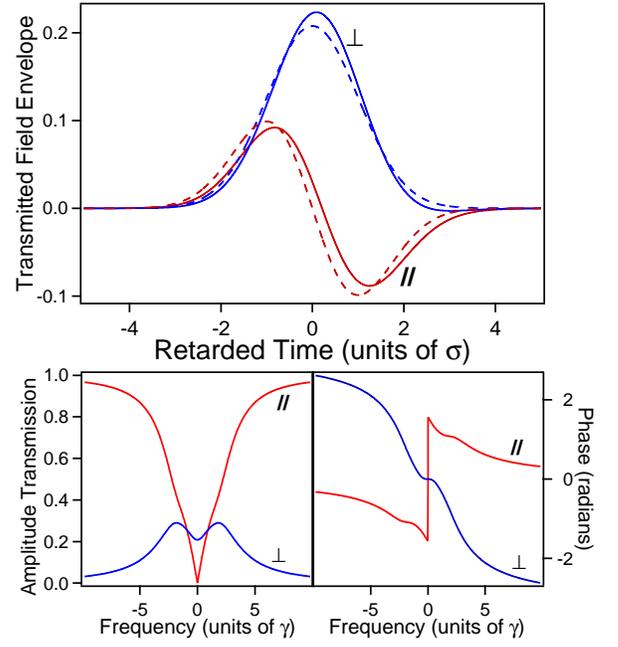} 
\par\end{centering}

\caption{Top: envelopes $y_{\parallel}(t)$ and $y_{\perp}(t)$
of the transmitted pulses for $x(t)=\exp{[}-t^{2}/(2\sigma^{2}){]}$.
The solid {(}dashed{)} lines are the exact envelopes obtained by fast
Fourier transforms {(}the approximate analytical envelopes $y_{\parallel}(t)$
given by Eq. (\ref{eq:16}) and $y_{\perp}(t)=e^{-\pi/2}x(t)${)}.
Parameters: $\Delta=\gamma$, $\theta=\pi/2$ ($\alpha\ell=2\pi$),
and $\sigma=2.0/\gamma$ . Bottom: corresponding amplitude transmission
and phase as a function of $\Omega$.\label{fig:Alegal2Pi}}
\end{figure}

Singular behaviors are obtained when $\theta=\left(2k-1\right)\pi/2$
for $H_{\parallel}(\Omega)$ and when $\theta=k\pi$ for $H_{\perp}(\Omega)$
where $k$ is a positive integer. The cw transmission $\left|H\left(0\right)\right|$
is then null, and the phase $\Phi\left(\Omega\right)$ displays a $\pi$
discontinuity at $\Omega=0$. Note that such phase singularities have
been clearly recognized in the analysis of the pioneering experiments
on non dispersive birefringent media \cite{so04,bru04}. As $\left|H\left(0\right)=0\right|$,
the area $\int_{-\infty}^{+\infty}y(t)\,dt$ of the envelope $y(t)$
of the transmitted pulse is also null \cite{re2}, its center of gravity
is not defined and the pulse distortion is considerable, as narrow
as the spectrum of the incident pulse may be. Figure \ref{fig:Alegal2Pi}
shows the results obtained for $\theta=\pi/2$. There is no problem for $H_{\perp}(\Omega)$
with $H_{\perp}(0)=e^{-\pi/2},$ $\tau_{g\perp}=0$ and,
if its spectrum is narrow enough, the incident pulse will propagate
without significant distortion at the velocity $c$. 
On the other hand $H_{\parallel}\left(0\right)=0$. The phase $\Phi_{\parallel}\left(\Omega\right)$
has the predicted discontinuity and the area of $y_{\parallel}(t)$
is actually null. Again if the spectrum of $x(t)$ is narrow enough,
$y_{\parallel}(t)$ can be calculated by using the power-series expansion
of $H_{\parallel}\left(\Omega\right)$ at the first order in $\Omega$,
which reads $H_{\parallel}\left(\Omega\right)\approx\theta e^{-\theta}\left(i\Omega/\gamma\right)$.
Passing in the time domain ($i\Omega\rightarrow d/dt$), we get 
\begin{equation}
y_{\parallel}(t)=\frac{\pi e^{-\pi/2}}{2\gamma}\frac{dx}{dt}=-\frac{\pi e^{-\pi/2}}{2\gamma\sigma}\left(\frac{t}{\sigma}\right)\exp\left(-\frac{t^{2}}{2\sigma^{2}}\right).\label{eq:16}
\end{equation}
For the value $\sigma=2/\gamma$ used in Fig. \ref{fig:Alegal2Pi},
$y_{\perp}(t)=e^{-\pi/2}x(t)$ and $y_{\parallel}(t)$ given
by Eq. (\ref{eq:16}) appear to be good approximations of the exact envelope
of the transmitted pulses.

When $\theta$ is close to the pathologic values considered in the
previous paragraph, the phase discontinuity is replaced by a rapid
variation around $\Omega=0$, and in agreement with the relations
$\gamma\tau_{g\parallel}=-\theta\tan\theta$ and $\gamma\tau_{g\perp}=\theta/\tan\theta$
, one of the group delays takes very large values. However, the fractional
delays (delays in units of $\sigma$) with moderate distortion that
can be obtained are not significantly larger than those evidenced
in the reference case $\theta=\pi/4$ (Fig.\ref{fig:AlegalPI}). Indeed,
the spectral domain where $d\Phi(\Omega)/d\Omega$ and $\left|H(\Omega)\right|$
do not vary too considerably is very narrow and large pulse durations
$\sigma$ are necessary to avoid significant distortion of the transmitted
pulse. In addition, the corresponding transmission is very low. When,
e.g., $\theta=\left(\pi/2\right)-\left(\pi/20\right)$ , we have $\gamma\tau_{g\parallel}=-19$
but a numerical simulation shows that $\sigma$ as large as $47/\gamma$
is required to obtain a distortion comparable to that of the reference
case. The fractional time-advancement of the transmitted pulse is
then about $40\%$ instead of $30\%$ in the reference case but this
is at the expense of a reduction of the pulse intensity by a factor exceeding
$300$! 
\begin{figure}[h]
\begin{centering}
\includegraphics[width=0.9\columnwidth]{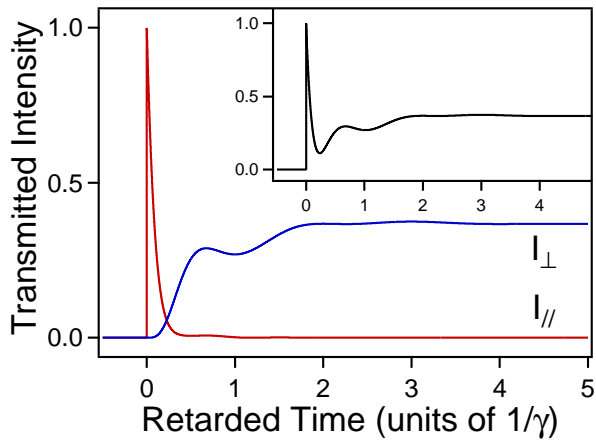} 
\par\end{centering}

\caption{Numerically determined intensity profiles of the transmitted
pulses when the incident pulse is step modulated with $x(t)=u_{H}(t)$.
Parameters: $\theta=\pi/2$ and $\Delta=\pi\gamma$ ($\alpha\ell=1+\pi^{2}$).
The optical precursor and the main field are, respectively, detected
in the parallel and perpendicular polarizations. The inset shows the
intensity profile obtained in the absence of the polarization beam
splitter.\label{fig:Precursor}}
\end{figure}

The observation of slow light and, particularly, of fast light requires
the use of incident pulses with ideally smooth envelope. We now consider
briefly the opposite case where the incident field is switched with a rise time infinitely short with respect to all the characteristic
times of the system but long compared to $1/\omega_{0}$, so that the
slowly varying envelope approximation remains valid \cite{al87}.
If the envelope $x(t)$ of the incident pulse is a Heaviside unit step function
$u_{H}(t)$, as currently considered in the study of optical precursors
\cite{che13}, that of the transmitted pulse reads 
\begin{equation}
y(t)=\int_{-\infty}^{t}h(t')dt',\label{eq:17}
\end{equation}
where $h(t)$ is the impulse response of the system, the inverse Fourier
transform of its transfer function $H(\Omega)$. Using standard results
of Laplace transforms \cite{ab72}, we get 
\begin{equation}
h_{\parallel}(t)=\delta(t)-\gamma\alpha\ell\frac{J_{1}\left(\sqrt{2\gamma\alpha\ell t}\right)}{\sqrt{2\gamma\alpha\ell t}}\cos\left(\Delta t\right) e^{-\gamma t}u_{H}(t),\label{eq:18}
\end{equation}
\begin{equation}
h_{\perp}(t)=\gamma\alpha\ell\frac{J_{1}\left(\sqrt{2\gamma\alpha\ell t}\right)}{\sqrt{2\gamma\alpha\ell t}}\sin\left(\Delta t\right) e^{-\gamma t}u_{H}(t).\label{eq:19}
\end{equation}
where $\delta\left(z\right)$ and $J_{1}\left(z\right)$ respectively
designate the Dirac delta function and the Bessel function of the
first kind and index $1$. Interesting features are obtained when
$\theta=\pi/2$ whatever $\Delta$ is. Figure  \ref{fig:Precursor} shows
the result obtained in this case for $\Delta=\pi\gamma$, a value
chosen so that the  precursor and steady state or main field have comparable
intensities and are not clearly distinguishable in the absence of
the PBS (see inset of Fig. \ref{fig:Precursor}).
A similar situation occurs in experiments involving a single absorption
line. See Ref. \cite{je06} and the related discussion in \cite{ma13}.
When the PBS is operating, our system separates the transmitted field
into two parts. The part detected in the parallel polarization can be
attributed to the precursor. Indeed, insofar as the field polarization
does not rotate instantaneously, its initial intensity equals $1$
but tends to zero as soon as the polarization rotates by $\pi/2$.
On the contrary, the part detected in the perpendicular polarization
starts from zero before asymptotically increasing to a steady value
and is nothing but the so-called main-field in the precursor theory.

To summarize, we have proposed a hybrid system associating the effects
of medium dispersion with the effects of polarization selection. At least
conceptually, this system is relatively simple. It enables one to
obtain simultaneous fast and slow light. Due to the non minimum phase shift
of its transfer function, it presents a great variety of behaviors only a few of which have been explored. Additionally it can also be used to separate optical precursor and main field when the incident wave
is step modulated.

This work has been partially supported by the Ministry of Higher Education
and Research, the Nord-Pas de Calais Regional Council and the European Regional
Development Fund (ERDF) through the Contrat de Projets État-Région
(CPER) 2015\textendash 2020, as well as by the Agence Nationale de
la Recherche through the LABEX CEMPI project (ANR-11-LABX-0007).


\begin{thebibliography}{10}
\bibitem{re00} If the magnetic field is not uniform, the
splitting of the Zeeman sublevels will suffer from an inhomogeneous
broadening which will be retrieved on the transitions $\sigma_{+}$
and $\sigma_{-}$. For the validity of the subsequent calculations,
it is necessary that this broadening remains small compared to the
homogeneous broadening.
\bibitem{re0} Note that this result holds even when the pulse is
strongly distorted. The time delay of its maximum then significantly
deviates from the group delay.
\bibitem{re1} Quite generally the group velocity reads $v_{g}=c\left[n(\omega)+dn(\omega)/d\omega\right]^{-1}$.
For the homogeneously broadened resonances in a dilute medium considered
here, $n(\omega)\approx1$ and $dn(\omega)/d\omega=0$ when $\omega$
is detuned from resonance by $\pm\gamma$. We have then $v_{g}\approx c$.
\end{thebibliography}
\end{document}